\documentclass[runningheads]{llncs}
\usepackage{graphicx}
\usepackage{multirow}
\usepackage[colorlinks,linkcolor=red, anchorcolor=blue, citecolor=blue, urlcolor=magenta]{hyperref}
\usepackage{framed,multirow}
\usepackage{threeparttable}
\usepackage[misc]{ifsym}
\usepackage[symbol]{footmisc}
\usepackage{amsmath}
\usepackage{amssymb}
\usepackage{algorithmic,algorithm}
\begin{document}
\title{Exploring Smoothness and Class-Separation for Semi-supervised Medical Image Segmentation}
\titlerunning{SS-Net for Semi-supervised Segmentation}
\author{Yicheng Wu \textsuperscript{1(\Letter)} \and Zhonghua Wu \inst{2} \and Qianyi Wu \inst{1} \and Zongyuan Ge\inst{3,4} \and Jianfei Cai\inst{1} }
\authorrunning{Yicheng Wu et al.}
\institute{\textsuperscript{1}Department of Data Science \& AI, Faculty of Information Technology, Monash University, Melbourne, VIC 3800, Australia
\\
\email{yicheng.wu@monash.edu}\\
\textsuperscript{2}School of Computer Science and Engineering, Nanyang Technological University, Singapore, 639798, Singapore \\
\textsuperscript{3}Monash-Airdoc Research, Monash University, Melbourne, VIC 3800, Australia\\
\textsuperscript{4}Monash Medical AI, Monash eResearch Centre, Melbourne, VIC 3800, Australia \\
}
\maketitle 
\begin{abstract}
Semi-supervised segmentation remains challenging in medical imaging since the amount of annotated medical data is often scarce and there are many blurred pixels near the adhesive edges or in the low-contrast regions. To address the issues, we advocate to firstly constrain the consistency of pixels with and without strong perturbations to apply a sufficient smoothness constraint and further encourage the class-level separation to exploit the low-entropy regularization for the model training.
Particularly, in this paper, we propose the SS-Net for semi-supervised medical image segmentation tasks, via exploring the pixel-level \textbf{S}moothness and inter-class \textbf{S}eparation at the same time. The pixel-level smoothness forces the model to generate invariant results under adversarial perturbations. Meanwhile, the inter-class separation encourages individual class features should approach their corresponding high-quality prototypes, in order to make each class distribution compact and separate different classes.
We evaluated our SS-Net against five recent methods on the public LA and ACDC datasets. Extensive experimental results under two semi-supervised settings demonstrate the superiority of our proposed SS-Net model, achieving new state-of-the-art (SOTA) performance on both datasets. The code is available at \url{https://github.com/ycwu1997/SS-Net}.
\end{abstract}

\section{Introduction}
Most of deep learning-based segmentation models rely on large-scale dense annotations to converge and generalize well. However, it is extremely expensive and labor-consuming to obtain adequate per-pixel labels for the model training. Such a heavy annotation cost has motivated the community to study the semi-supervised segmentation methods \cite{ssl,wu2019keypoint,liu2021few}, aiming to train a model with few labeled data and abundant unlabeled data while still achieving a satisfactory segmentation performance.

Existing semi-supervised segmentation methods are usually based on the two assumptions: smoothness and low entropy. The smoothness assumption encourages the model to generate invariant results under small perturbations at the data level \cite{fixmatch,strong,vat,transformation},  the feature level \cite{cct,dcc,pair,wu2021learning} and the model level \cite{cotraining,dtc,urpc,uamt,sassnet}. Its success implies that the local distributional smoothness (LDS) is crucial to leverage abundant unlabeled data for the model training. The low-entropy assumption further constrains that the decision boundary should lie in the low-density regions \cite{pseudo,metapseudolabels,mcnet,mc+}. In other words, the semi-supervised models are expected to output highly confident predictions even without the supervision of corresponding labels, aiming at the inter-class separation.

\begin{figure*}[t]
\centering
\includegraphics[width=1\textwidth]{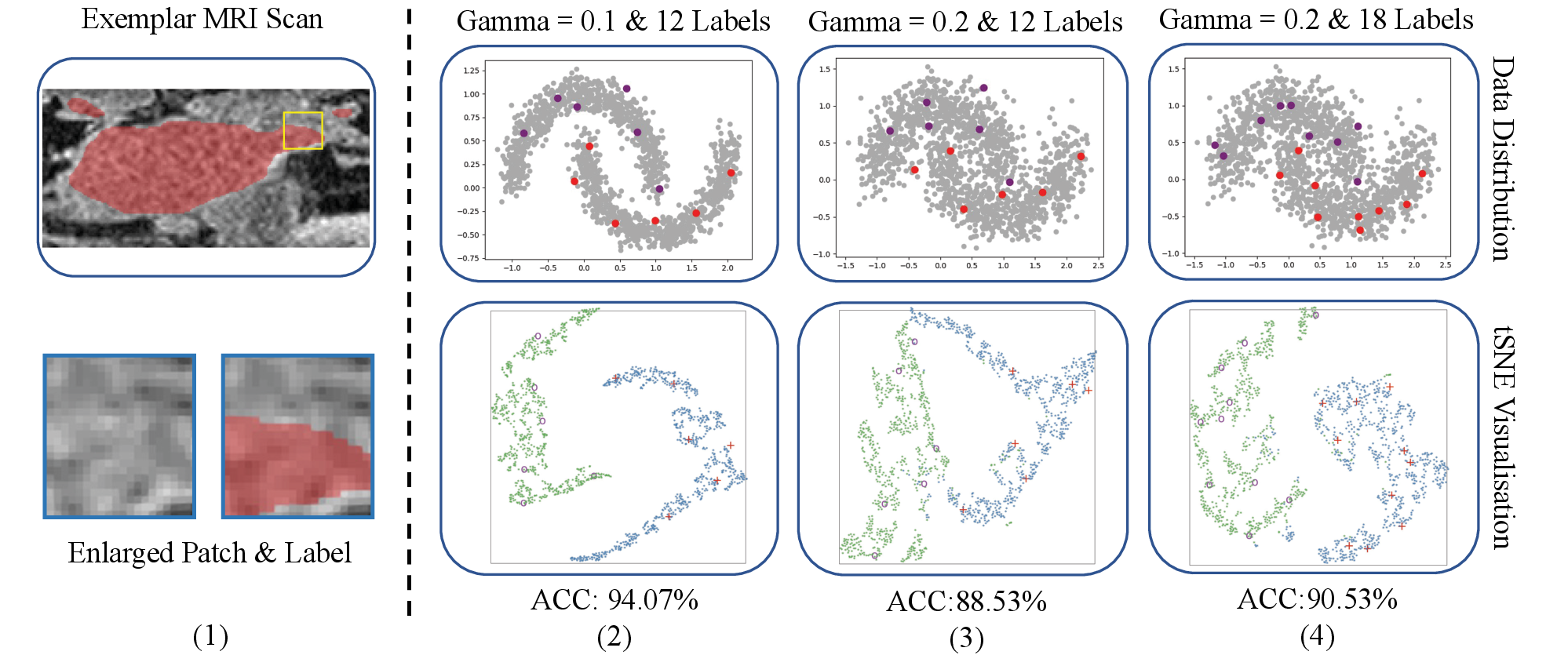}
\caption{\label{motivation} Exemplar MRI scan (Top Left) and illustrations of three two-moon datasets (Top Right), equipped with the enlarged patch/label (Bottom Left) and their tSNE \cite{tsne} visualized features (Bottom Right). The gamma is used to control the class dispersion.}
\end{figure*}
Despite the progress, the semi-supervised segmentation in the medical imaging area remains challenging due to two factors: fewer labels and blurred targets.
As illustrated in Fig.~\ref{motivation}, there are many ambiguous pixels near the adhesive edges or low-contrast regions and the amount of labeled data is usually limited. Such concomitant challenges may lead to poor performance of the existing models \cite{uamt,sassnet,dtc,urpc,mcnet}. 
We further use the synthetic two-moon dataset as a toy example to illustrate this scenario. Specifically, we trained a four-layer MLP in a supervised manner and then used the tSNE tool \cite{tsne} to visualize the deep features. We can see that, with fewer labels and more blurred data (larger gamma), the performance drops significantly and the model cannot distinguish different classes well since the feature manifolds of different classes are inter-connected, see the 3rd column of Fig.~\ref{motivation}, where the low-entropy assumption is violated.

To alleviate the problems, we advocate: (1) the \textit{strong perturbations} are needed to sufficiently regularize the large amounts of unlabeled medical data \cite{strong};  (2) the \textit{class-level separation} is also suggested to pursue the decision boundary in low-density regions. Therefore, in this paper, we propose the SS-Net, to explore the pixel-level \textbf{S}moothness and inter-class \textbf{S}eparation at the same time for semi-supervised medical image segmentation. Specifically, our SS-Net has two novel designs. First, inspired by the virtually adversarial training (VAT) model \cite{vat}, we adopt the adversarial noises as strong perturbations to enforce a sufficient smoothness constraint. Second, to encourage the inter-class separation, we select a set of high-quality features from the labeled data as the prototypes and force other features to approach their corresponding prototypes, making each class distribution compact and pushing away different classes. We evaluated our SS-Net on the public LA and ACDC datasets \cite{la,acdc}. Extensive experimental results demonstrate our model is able to achieve significant performance gains.

Overall, our contributions are three-fold:
(1) we point out two challenges, \textit{i.e.}, fewer labels and blurred targets, for the semi-supervised medical image segmentation and show our key insight that it is crucial to employ the strong perturbations to sufficiently constrain the pixel-level smoothness while at the same time encouraging the inter-class separation, enabling the model to produce low-entropy predictions;
(2) in our SS-Net, we introduce the adversarial noises as strong perturbations. To our knowledge, it is one of the first to apply this technique to perturb medical data for semi-supervised tasks. Then, the inter-class separation is encouraged via shrinking each class distribution, which leads to better performance and complements the pixel-level smoothness;
(3) via utilizing both techniques, our SS-Net outperforms five recent semi-supervised methods and sets the new state of the art on the LA and ACDC datasets.
\section{Method}
\begin{figure*}[htb]
\centering
\includegraphics[width=1\textwidth]{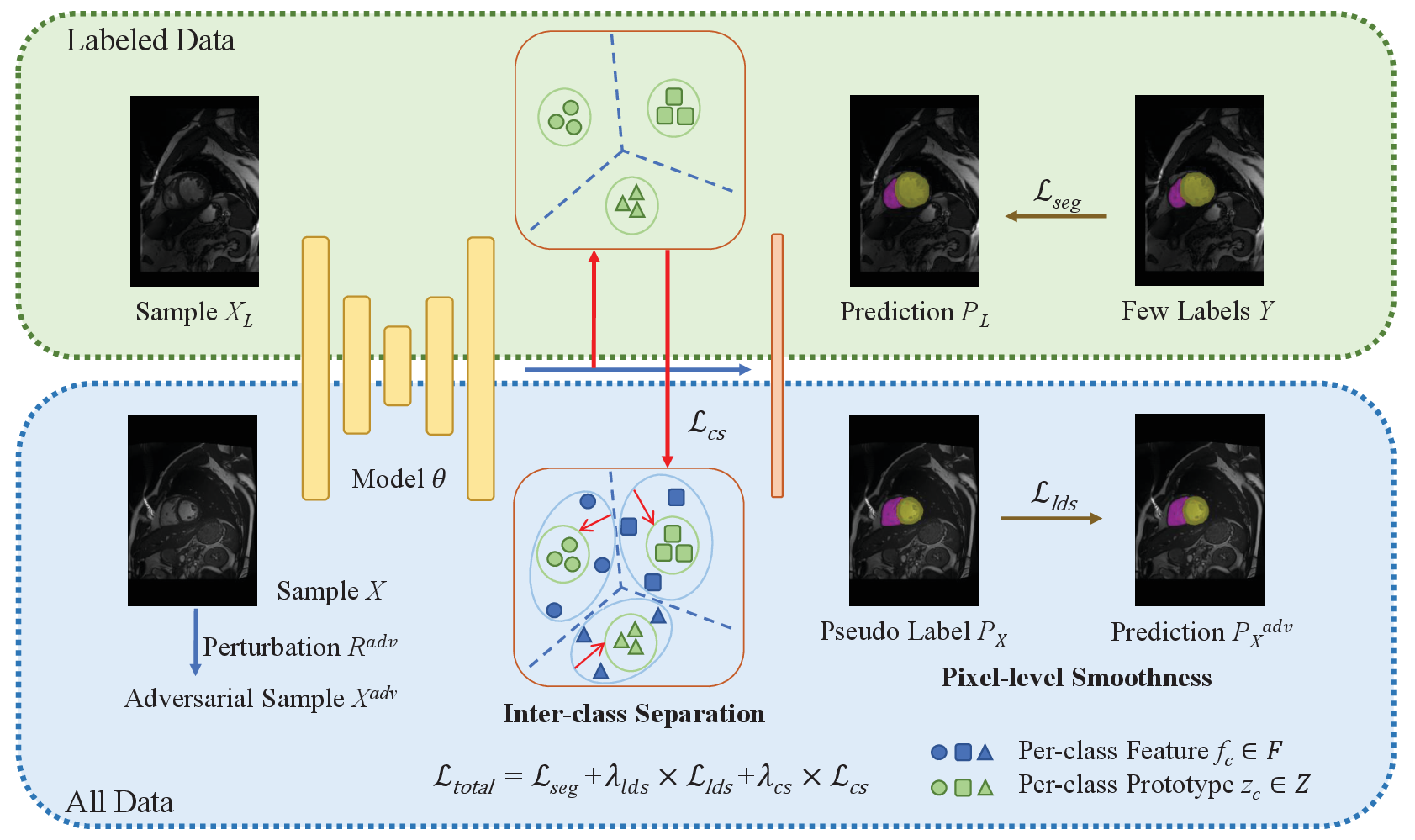}
\caption{\label{pipeline}Pipeline of our proposed SS-Net. The pixel-level smoothness is constrained at the pixel level via applying adversarial noises as strong perturbations while the inter-class separation is performed at the feature level by a prototype-based strategy.}
\end{figure*}
Fig.~\ref{pipeline} shows the overall pipeline of our SS-Net. We propose two designs to encourage the pixel-level smoothness and the inter-class separation, respectively. First, the pixel-level smoothness is enforced via applying a consistency constraint between an original image $x \in X$ and its perturbed sample with the per-pixel adversarial noises. Second, we compute a set of feature prototypes $Z$ from the labeled data $X_L$, and then encourage high dimensional features $F$ to be close to the prototypes $Z$ so as to separate different classes in the feature space. We now delve into the details.
\subsection{Pixel-level Smoothness}
It is nowadays widely recognized that the LDS is critical for semi-supervised learning \cite{ssl}. This kind of regularization can be formulated as:
\begin{equation}\label{lds}
    \mathcal{LDS}(x; \theta) = D\left[\hat{y}, p(x+r)\right], \|r\|\leq\epsilon
\end{equation}
where $D$ is used to compute the discrepancy between the prediction of a perturbed sample and its true label $\hat{y} \in Y$, and $\epsilon$ controls the magnitude of the enforced perturbation $r$. Since adequate true labels are not available in the semi-supervised scenario, $\hat{y}$ is usually set as the pseudo label $p(\hat{y}|x)$. Essentially, LDS regularizes the model to be robust or consistent with small perturbations of data.

Meanwhile, to apply strong perturbations, following the VAT model \cite{vat}, we use the gradient $g$ as the direction of $r^{adv}$ to perturb the original sample $x$, which is at the pixel level and can be estimated as:
\begin{equation}
\label{gradient}
\begin{aligned}
&g = \nabla_{r^{ini}} D\left[p(\hat{y}|x), p(y|x+r^{ini})\right] \\
&r^{adv} = \epsilon \times g / \|g\|_2,
\end{aligned}
\end{equation}
where $r^{ini}$ can be set as a random noise vector, $g$ denotes the fastest-changing direction at the measurement of $D$, and $r^{adv} \in R^{adv}$ is the corresponding adversarial noise vector. Note that, in the original VAT model \cite{vat}, $D$ is adopted as the K-L Divergence. However, through experiments, we found that the K-L Divergence might not be a suitable one for segmentation tasks. Therefore, we utilize the Dice loss as $D$ to generate adversarial noises and the LDS loss becomes
\begin{equation}
\label{losslds}
\begin{aligned}
\mathcal{L}_{lds} =\frac{1}{C} \sum_{c=1}^C [1-\frac{2\| p(\hat{y_c}|x) \cap  p(y_c|x+r^{adv})\|}{ \|p(\hat{y_c}|x)\| + \|p(y_c|x+r^{adv})\|}]
\end{aligned}
\end{equation}
where $C$ is the number of classes. $p(\hat{y_c}|x)$ and $p(y_c|x+r^{adv})$ denote the predictions of $x$ with and without strong perturbations in the $c$-th class, respectively.

In this way, $g$ can be efficiently computed via the back-propagation scheme. Compared to random noises, such adversarial noises can provide a stronger smoothness constraint to facilitate the model training \cite{vat}. 

\subsection{Inter-class Separation}
When segmenting ambiguous targets, only enforcing LDS is insufficient since the blurred pixels near the decision boundary could be easily assigned to uncertain labels, which can confuse the model training. Therefore, to complement LDS, we further encourage the inter-class separation in the feature space. Compared with directly applying entropy minimization to the results, this feature-level constraint is more effective for semi-supervised image segmentation \cite{cct}.

Therefore, we employ a prototype-based strategy \cite{classa,xu2022all} to disconnect the feature manifolds of different classes, which can reduce the computational costs. Specifically, we first use non-linear projectors to obtain the projected features $F=F_l \cup F_u$. Then, a subset of labeled features $F_l$ is selected according to their correct predictions in the target categories. Next, we sort these candidate features via the ranking scores generated by $C$ attention modules, and the top-K highest-scoring features are finally adopted as the high-quality prototypes $Z$.

Afterward, we leverage current predictions to group individual class features $f_c\in F$ and force them to approach their corresponding prototype $z_c\in Z$, aiming to shrink the intra-class distribution. We use the cosine similarity to compute the distance between $z_c$ and $f_c$ with the loss $\mathcal{L}_{cs}$ defined as
\begin{equation}
\label{losSS}
\begin{aligned}
\mathcal{L}_{cs} = \frac{1}{C} \frac{1}{N} \frac{1}{M} \sum_{c=1}^{C} \sum_{i=1}^{N} \sum_{j=1}^{M}w_{ij}(1-\frac{\langle z_c^i, \ f_c^j\rangle}{\|z_c^i\|_2 \cdot \|f_c^j\|_2})
\end{aligned}
\end{equation}
where $w_{ij}$ is the weight for normalization as \cite{classa}, and $N$ or $M$ respectively denotes the number of prototypes or projected features in the $c$-th class. Here, $\mathcal{L}_{cs}$ can align the labeled/unlabeled features and make each class distribution compact, resulting in a good separation of different classes in the feature space. 

Finally, the overall loss is a weighted sum of the segmentation loss $\mathcal{L}_{seg}$ and the other two losses:
\begin{equation}
\label{totol}
\begin{aligned}
\mathcal{L}_{total} = \mathcal{L}_{seg} + \lambda_{lds} \times \mathcal{L}_{lds} + \lambda_{cs} \times \mathcal{L}_{cs} 
\end{aligned}
\end{equation}
where $\lambda_{lds}$ and $\lambda_{cs}$ are the corresponding weights to balance the losses. Note that, $\mathcal{L}_{seg}$ is a Dice loss, which is applied for the few labeled data. The other two losses are applied for all data to regularize the model training.
\section{Experiment and Results}
\subsection{Dataset}
We evaluate the proposed SS-Net on the LA\footnote{\url{http://atriaseg2018.cardiacatlas.org}} dataset \cite{la} and the ACDC\footnote{\url{https://www.creatis.insa-lyon.fr/Challenge/acdc/databases.html}} dataset \cite{acdc}. The LA dataset consists of 100 gadolinium-enhanced MRI scans, with a fixed split\footnote{\url{https://github.com/yulequan/UA-MT/tree/master/data}} of 80 samples for training and 20 samples for validation. We report the performance on the validation set for fair comparisons as \cite{uamt,sassnet,dtc,mcnet}. On the ACDC dataset, the data split\footnote{\url{https://github.com/HiLab-git/SSL4MIS/tree/master/data/ACDC}} is also fixed with 70, 10, and 20 patients' scans for training, validation, and testing, respectively. All experiments follow the identical setting for fair comparisons and we consider the challenging semi-supervised settings to verify our model, where 5\% and 10\% labels are used and the rest in the training set are regarded as unlabeled data.
\subsection{Implementation Details}
Following the public methods \cite{uamt,sassnet,dtc,mcnet,urpc} on both datasets, all inputs were normalized as zero mean and unit variance. We used the rotation and flip operations to augment data and trained our model via a SGD optimizer with a learning rate 0.01. The loss weights $\lambda_{lds}$ and $\lambda_{cs}$ were set as an iteration-dependent warming-up function \cite{rampup} and $w_{ij}$ was obtained by normalizing the learnable attention weights as \cite{classa}. We updated each class-level feature prototype of size $32\times32$ in an online fashion.
On LA, $\epsilon$ was set as 10 and we chose the V-Net \cite{vnet} as the backbone. For training, we randomly cropped $112\times112\times80$ patches and the batch size was 4, containing two labeled patches and two unlabeled patches. We trained the model via 15K iterations. For testing, we employed a fixed stride ($18\times18\times4$) to extract patches and the entire predictions were recomposed from patch-based outputs. On ACDC, we set $\epsilon$ as 6 and chose the U-Net model \cite{unet} as the backbone to process 2D patches of size $256\times256$. The batch size was 24 and the total training iterations were 30K. All experiments in this paper were conducted on the same environments with fixed random seeds (Hardware: Single NVIDIA Tesla V100 GPU; Software: PyTorch 1.8.0+cu111, and Python 3.8.10).
\subsection{Results}
\begin{figure*}[!htb]
\centering
\includegraphics[width=1\textwidth]{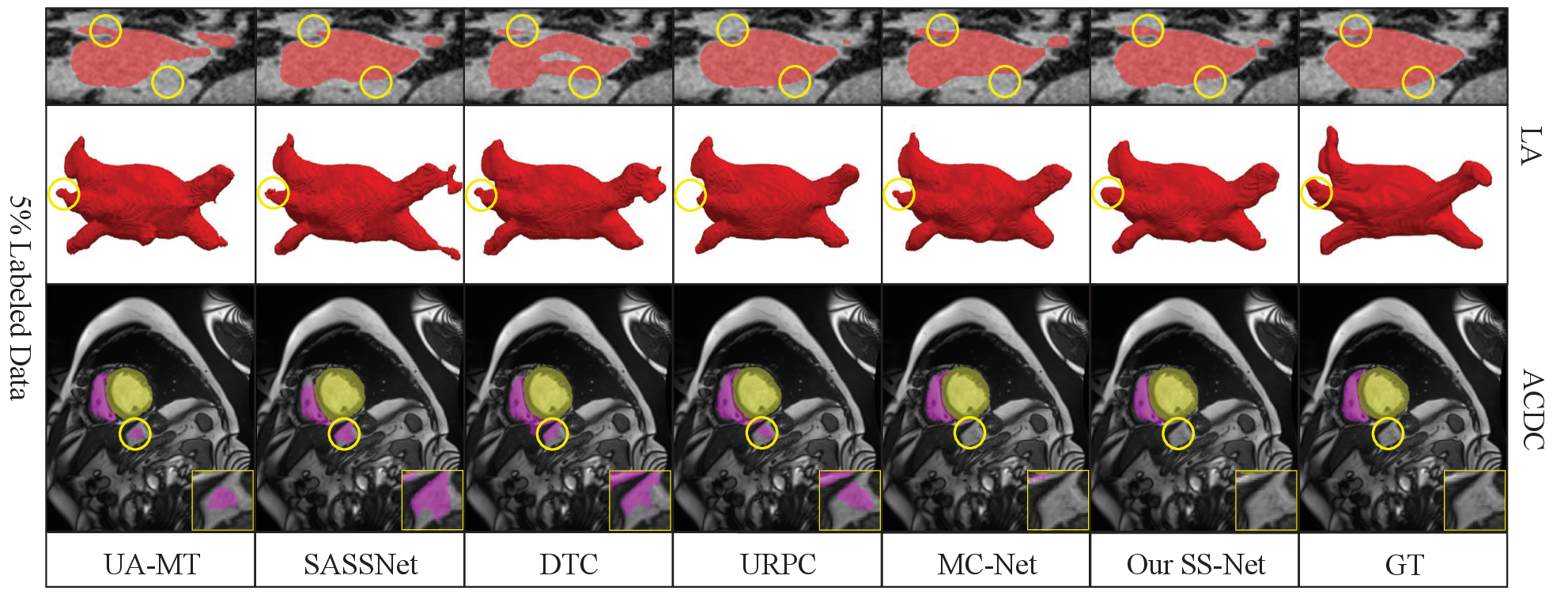}
\caption{\label{results}Exemplar results of several semi-supervised segmentation methods with 5\% labeled data training and corresponding ground truth (GT) on the LA dataset (Top two rows) and ACDC dataset (Bottom row).}
\end{figure*}
\subsubsection{Performance on the LA dataset}In Table~\ref{resultla}, we use the metrics of Dice, Jaccard, 95\% Hausdorff Distance (95HD), and Average Surface Distance (ASD) to evaluate the results. It reveals that: (1) compared to the lower bounds, \textit{i.e.}, only with 5\%/10\% labeled data training, our proposed SS-Net can effectively leverage the unlabeled data and produce impressive performance gains of all metrics; (2) when trained with fewer labels, \textit{i.e.}, 5\%, our SS-Net significantly outperforms other models on the LA dataset. It indicates enforcing the adversarial perturbations is useful to sufficiently regularize the unlabeled data when the amount of labeled data is limited. Furthermore, we post-processed all the results via selecting the largest connected component as \cite{sassnet} for fair comparisons. Note that we did not enforce any boundary constraint to train our model and our model naturally achieves satisfied shape-related performance on the LA dataset. The visualized results in Fig.~\ref{results} indicates that our SS-Net can detect most organ details, especially for the blurred edges and thin branches (highlighted by yellow circles), which are critical attributes for clinical applications.
{
\begin{table*}[!htb]
	\centering
	\caption{Comparisons with five recent methods on the LA dataset. All results were re-produced as \cite{uamt,sassnet,dtc,urpc,mcnet} in the identical experimental setting for fair comparisons.}
	\label{resultla}
    \begin{threeparttable}
	\resizebox{\textwidth}{!}{
	\begin{tabular}{c|cc|cccc|cc}
		\hline 
		\hline
		\multirow{2}{*}{Method}&\multicolumn{2}{c}{\# Scans used}&\multicolumn{4}{|c}{Metrics}&\multicolumn{2}{|c}{Complexity}\\
		\cline{2-9}
		&Labeled&Unlabeled &Dice(\%)$\uparrow$ &Jaccard(\%)$\uparrow$&95HD(voxel)$\downarrow$&ASD(voxel)$\downarrow$&Para.(M)&MACs(G)\\
		\hline
		V-Net & 4(5\%) &0 &52.55 &39.60 &47.05 &9.87 &9.44 &47.02\\
		V-Net &8(10\%) &0 &82.74 &71.72 &13.35 &3.26 &9.44 &47.02\\
		V-Net &80(All) &0 &91.47 &84.36 &5.48 &1.51 &9.44 &47.02\\
		\hline
		UA-MT \cite{uamt} (MICCAI'19) & \multirow{6}{*}{4 (5\%)} &\multirow{6}{*}{76 (95\%)} &82.26 &70.98 &13.71 &3.82 &9.44 &47.02\\
		SASSNet \cite{sassnet} (MICCAI'20) &  & &81.60 &69.63 &16.16 &3.58 &9.44 &47.05\\
		DTC \cite{dtc} (AAAI'21) &  & &81.25 &69.33 &14.90 &3.99 &9.44 &47.05\\
		URPC \cite{urpc} (MICCAI'21) &  & &82.48 &71.35 &14.65 &3.65 &5.88&69.43\\
		MC-Net \cite{mcnet}	(MICCAI'21) &  & &83.59 & 72.36 &14.07 &2.70 &12.35 &95.15\\
		SS-Net (Ours) &  & &\textbf{86.33} &\textbf{76.15} &\textbf{9.97} &\textbf{2.31} &9.46 &47.17\\
		\hline
		UA-MT \cite{uamt} (MICCAI'19) & \multirow{6}{*}{8 (10\%)} &\multirow{6}{*}{72 (90\%)} &87.79 &78.39 &8.68 &2.12 &9.44 &47.02\\
		SASSNet \cite{sassnet} (MICCAI'20) &  & &87.54 &78.05 &9.84 &2.59 &9.44 &47.05\\
		DTC \cite{dtc} (AAAI'21) &  & &87.51 &78.17 &8.23 &2.36 &9.44 &47.05\\
		URPC \cite{urpc} (MICCAI'21) &  & &86.92 &77.03 &11.13 &2.28 &5.88&69.43\\
		MC-Net \cite{mcnet}	(MICCAI'21) &  & &87.62 &78.25 &10.03 & \textbf{1.82} &12.35 &95.15\\
		SS-Net (Ours) &  & &\textbf{88.55} &\textbf{79.62} &\textbf{7.49} &1.90 &9.46 &47.17\\
		\hline
		\hline
	\end{tabular}}
    \end{threeparttable}
\end{table*}
}
{
\begin{table*}[!htb]
	\centering
	\caption{Comparisons with five recent methods on the ACDC dataset. All results were re-produced as \cite{uamt,sassnet,dtc,urpc,mcnet} in the identical experimental setting for fair comparisons.}
	\label{resultacdc}
    \begin{threeparttable}
	\resizebox{\textwidth}{!}{
	\begin{tabular}{c|cc|cccc|cc}
		\hline 
		\hline
		\multirow{2}{*}{Method}&\multicolumn{2}{c}{\# Scans used}&\multicolumn{4}{|c}{Metrics}&\multicolumn{2}{|c}{Complexity}\\
		\cline{2-9}
		&Labeled&Unlabeled &Dice(\%)$\uparrow$ &Jaccard(\%)$\uparrow$&95HD(voxel)$\downarrow$&ASD(voxel)$\downarrow$&Para.(M)&MACs(G)\\
		\hline
		U-Net & 3 (5\%) &0 &47.83 &37.01 &31.16 &12.62 &1.81&2.99\\
		U-Net & 7 (10\%) &0 &79.41 &68.11 &9.35 &2.70 &1.81&2.99\\
		U-Net & 70 (All) &0 &91.44 &84.59 &4.30 &0.99 &1.81&2.99\\
		\hline
		UA-MT \cite{uamt} (MICCAI'19)& \multirow{6}{*}{3 (5\%)} &\multirow{6}{*}{67 (95\%)} &46.04 &35.97 &20.08 &7.75 &1.81&2.99\\
		SASSNet \cite{sassnet} (MICCAI'20)  &  & &57.77 &46.14 &20.05 &6.06 &1.81&3.02\\
		DTC \cite{dtc} (AAAI'21)  &  &&56.90 &45.67 &23.36 &7.39 &1.81&3.02\\
		URPC \cite{urpc} (MICCAI'21) &  & &55.87 &44.64 &13.60 &3.74 &1.83&3.02\\
		MC-Net \cite{mcnet} (MICCAI'21)  & & &62.85 &52.29 &7.62 &2.33 &2.58&5.39\\
		SS-Net (Ours) & &  &\textbf{65.82} &\textbf{55.38} &\textbf{6.67} &\textbf{2.28} &1.83 &2.99\\
		\hline
		UA-MT \cite{uamt} (MICCAI'19)& \multirow{6}{*}{7 (10\%)} &\multirow{6}{*}{63 (90\%)} &81.65 &70.64&6.88 &2.02 &1.81&2.99\\
		SASSNet \cite{sassnet} (MICCAI'20)  &  & &84.50 &74.34 &5.42 &1.86 &1.81&3.02\\
		DTC \cite{dtc} (AAAI'21)  &  & &84.29 &73.92 &12.81 &4.01 &1.81&3.02\\
		URPC \cite{urpc} (MICCAI'21) &  & &83.10 &72.41 &\textbf{4.84} &1.53 &1.83&3.02\\
		MC-Net \cite{mcnet} (MICCAI'21)  &  & &86.44 &77.04 &5.50 &1.84 &2.58&5.39\\
		SS-Net (Ours) & & &\textbf{86.78} &\textbf{77.67} &6.07 &\textbf{1.40} &1.83&2.99\\
		\hline
		\hline
	\end{tabular}}
    \end{threeparttable}
\end{table*}
}
\subsubsection{Performance on the ACDC dataset}
Table~\ref{resultacdc} gives the averaged performance of three-class segmentation including the myocardium, left and right ventricles on the ACDC dataset. We can see that, with 5\% labeled data training, the performance of UA-MT model \cite{uamt} decreases significantly and is even worse than the lower bound, \textit{i.e.}, 46.04\% vs. 47.83\% in Dice. Since \cite{uamt} filters highly uncertain regions during training, such a performance drop suggests that it is needed to make full use of the ambiguous pixels, especially in the regime with extremely scarce labels. On the contrary, as Table~\ref{resultacdc} shows, our SS-Net surpasses all other methods and achieves the best segmentation performance. The bottom row of Fig.~\ref{results} also shows that our model can output good segmentation results and effectively eliminate most of the false-positive predictions on ACDC.
{
\begin{table*}[!htb]
	\centering
	\caption{Ablation studies on the LA dataset.}
	\label{ablation}
    \begin{threeparttable}
	\resizebox{\textwidth}{!}{
	\begin{tabular}{cc|ccc|cccc}
		\hline 
		\hline
		\multicolumn{2}{c}{\# Scans used}&\multicolumn{3}{|c}{Loss}&\multicolumn{4}{|c}{Metrics}\\
		\cline{1-9}
		Labeled&Unlabeled&$\mathcal{L}_{seg}$&$\mathcal{L}_{lds}$ &$\mathcal{L}_{cs}$ &Dice(\%)$\uparrow$ &Jaccard(\%)$\uparrow$&95HD(voxel)$\downarrow$&ASD(voxel)$\downarrow$\\
		\hline
		\multirow{5}{*}{4 (5\%)} &0 &\checkmark &   & &52.55 &39.60 &47.05 &9.87\\
		&76 (95\%) &\checkmark &\checkmark* & &82.27 &70.46 &13.82 &3.48 \\
	    &76 (95\%) &\checkmark &\checkmark & &84.31 &73.50 &12.91 &3.42 \\
		&76 (95\%) &\checkmark & &\checkmark &84.10 &73.36 &11.85 &3.36 \\
	    &76 (95\%) &\checkmark &\checkmark &\checkmark &\textbf{86.33} &\textbf{76.15} &\textbf{9.97} &\textbf{2.31} \\
		\hline
		\multirow{5}{*}{8 (10\%)}& 0 &\checkmark &  & &82.74 &71.72 &13.35 &3.26\\
		&72 (90\%) &\checkmark &\checkmark* & &87.48 &77.98 &8.95 &2.11 \\
		&72 (90\%) &\checkmark &\checkmark & &87.50 &77.98 &9.44 &2.08 \\
		&72 (90\%) &\checkmark & &\checkmark &87.68 &78.23 &8.84 &2.12 \\
		&72 (90\%) &\checkmark &\checkmark &\checkmark &\textbf{88.55} &\textbf{79.62} &\textbf{7.49} &\textbf{1.90} \\
		\hline
		\hline
	\end{tabular}}
    \begin{tablenotes}
    \footnotesize
    \item[*] We adopt the K-L Divergence as $D$, following the traditional VAT model \cite{vat}.
    \end{tablenotes}
    \end{threeparttable}
\end{table*}
}
\subsubsection{Ablation Study}
We conducted ablation studies on LA to show the effects of individual components. Table~\ref{ablation} indicates that either using $\mathcal{L}_{lds}$ to pursue the pixel-level smoothness or applying $\mathcal{L}_{cs}$ to encourage the inter-class separation is effective to improve the semi-supervised segmentation performance. Table~\ref{ablation} also gives the results of using K-L Divergence to estimate the adversarial noises as \cite{vat}, whose results suggest that adopting the Dice loss as $D$ can achieve better performance (\textit{e.g.}, 84.31\% vs. 82.27\% in Dice with 5\% labeled training data).
\section{Conclusion}
In this paper, we have presented the SS-Net for semi-supervised medical image segmentation. Given that fewer labels and blurred targets in the medical domain, our key idea is that it is important to simultaneously apply the adversarial noises for a sufficient smoothness constraint and shrink each class distribution to separate different classes, which can effectively exploit the unlabeled training data. The experimental results on the LA and ACDC datasets have demonstrated our proposed SS-Net outperforms other methods and achieves a superior performance for semi-supervised medical image segmentation. Future work will include the adaptive selections of the perturbation magnitude and the prototype size.
\subsubsection{Acknowledgements}
This work was supported by Monash FIT Start-up Grant. We also appreciate the efforts to collect and share the LA and ACDC datasets \cite{la,acdc} and several public repositories \cite{uamt,dtc,sassnet,mcnet,urpc}.
\bibliographystyle{splncs04}
\bibliography{paper}

\newpage
\section{Appendix}
\begin{figure*}[!htb]
\centering
\includegraphics[width=1\textwidth]{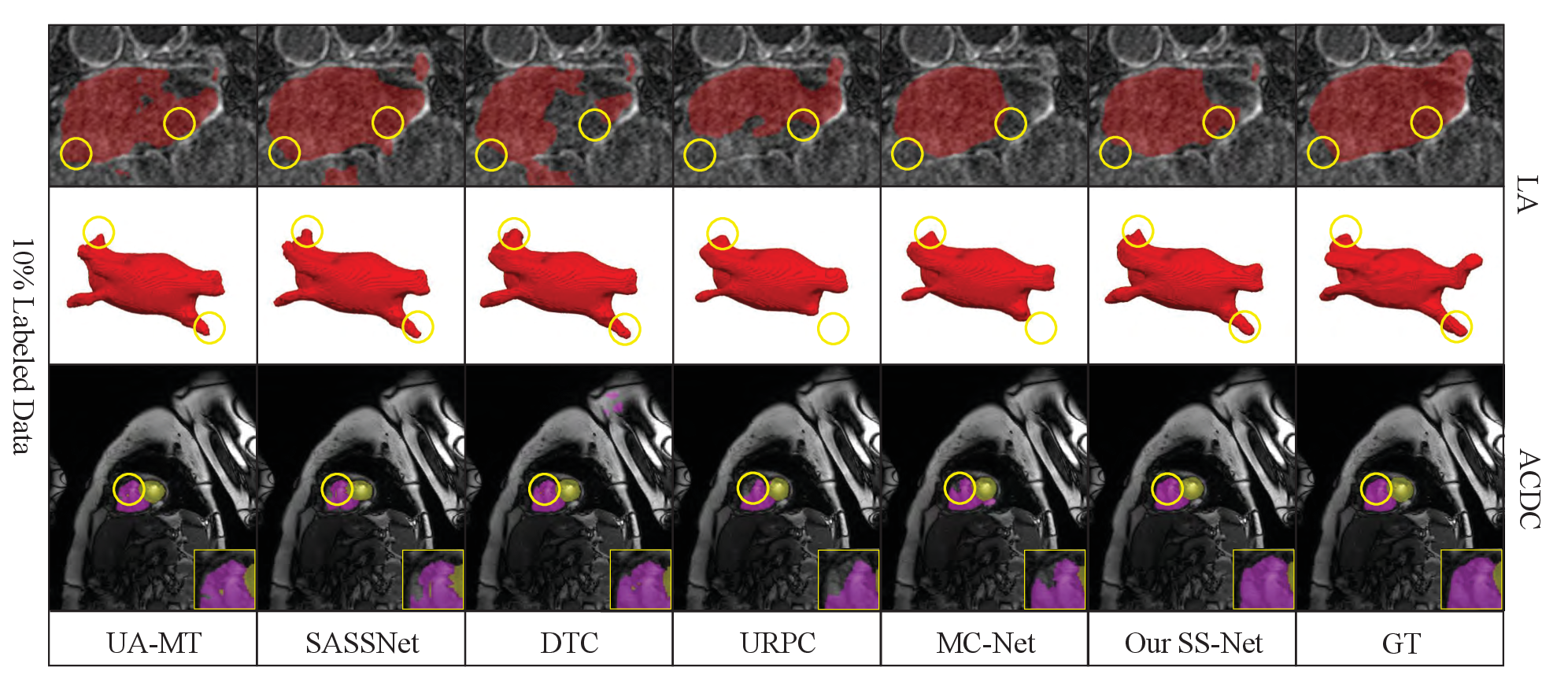}
\caption{\label{results_supp}Exemplar results of several semi-supervised segmentation methods with 10\% labeled data training and corresponding ground truth (GT) on the LA dataset (Top two rows) and ACDC dataset (Bottom row).}
\end{figure*}

\begin{algorithm}[!htb]
\caption{\label{algo:lds} Constraint for the pixel-level smoothness}
    \textbf{Input:}{ A batch of training sample $x$, perturbation magnitude $\epsilon$.}\\
    \textbf{Output:}{ $\mathcal{L}_{lds}$ for the model training.}
\begin{enumerate}
\item Generating the initial random noises $r^{ini}$ as the shape of $x$.
\item Adopting the Dice loss as $D$. \\
$D(a,b) = 1 - \frac{2 \|a\cap b\|}{\|a\| + \|b\|}$
\item Estimating the gradient of $D$ with respective to $r^{ini}$. \\
$g \leftarrow \nabla_{r^{ini}} D\left[p(\hat{y}|x), p(y|x+r^{ini})\right]$
\item Normalizing the g to generate adversarial perturbations $r^{adv}$.\\
$r^{adv} \leftarrow \epsilon \cdot g/\|g\|_2$
\item Obatining the adversarial examples $x^{adv}$. \\
$x^{adv} = x + r^{adv}$
\item Computing $\mathcal{L}_{lds}$ for the model training. \\
$\mathcal{L}_{lds} = D\left[p(\hat{y}|x), p(y|x^{adv})\right]$
\end{enumerate}
\end{algorithm}

\begin{figure*}[!htb]
\centering
\includegraphics[width=1\textwidth]{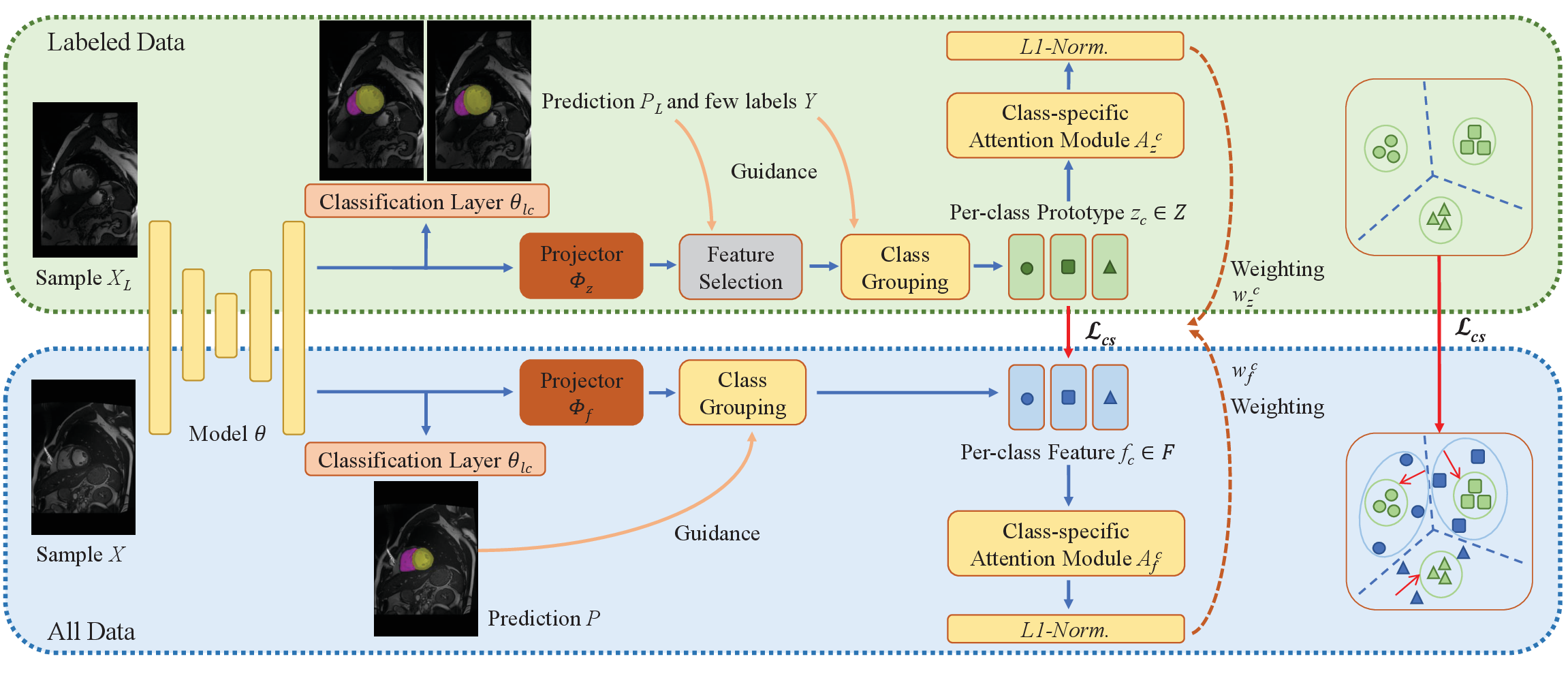}
\caption{\label{pipeline_cs}Pipeline of applying the $\mathcal{L}_{cs}$ for the model training as [1].}
\end{figure*}
\begin{algorithm}[!htb]
\caption{\label{algo:cs} Constraint for the inter-class separation}
    \textbf{Input:}{ Labeled and unlabeled deep features $Df_l$ and $Df_u$, prototype number $K$, last classification layer $\theta_{lc}$ with a SoftMax function, few labels $Y$, projectors $\Phi_{z}$ and $\Phi_{f}$, and class-specific attention modules $A_z^c$ and $A_f^c$. }\\
    \textbf{Output:}{ $\mathcal{L}_{cs}$ for the model training.}
\begin{enumerate}
\item Using $\Phi_{z}$ or $\Phi_{f}$ to respectively project the $Df_l$ or $Df$ ($Df = Df_l \cup Df_u$). \\
$F_l = \Phi_z(Df_l)$, $F = \Phi_f(Df)$
\item Selecting the correct labeled features as the prototype candidates $Z_{candidate}$ \\
(ignore the background category). \\
$Z_{candidate} = \{F_l| argMax(\theta_{lc}(Df_l)) == Y\ \&\& \  argMax(\theta_{lc}(Df_l)))>0\}$
\item Obtaining the learnable attention values of $P_{candidate}$ as the ranking scores. \\
$S_l = A_z(Z_{candidate})$
\item Selecting the top-K-scoring feature vectors as the Prototype $Z$. \\
$Z=Z_{candidate}[sort(S_l).indices,:][:K,:]$
\item Grouping each class according to the true labels $Y$ and pseudo labels, i.e., the current prediction $\theta_{lc}(Df_u)$. \\
$z_c \in Z$, $f_c \in F$\\
where $z_c$ and $f_c$ denote the prototype and the projected feature of $c$-th class.
\item Normalizing the attention values to weight the $z_c$ and $f_c$. \\
$w_z^c = L1_{norm}\{A_z^c(Z)\}$, $w_f^c = L1_{norm}\{A_f^c(F)\}$
\item Adopting the cosine similarity to compute the distance between $Z$ and $F$.\\
$CS(z,f) = 1-\frac{\langle z, \ f\rangle}{\|z\|_2 \cdot \|f\|_2}$
\item Computing $\mathcal{L}_{cs}$ for the model training. \\
$\mathcal{L}_{cs} = \frac{1}{C} \frac{1}{N} \frac{1}{M} \sum_{c=1}^{C} \sum_{i=1}^{N} \sum_{j=1}^{M}w_{zi}^cw_{fj}^c CS(z_c^i, f_c^j) $ \\
where $N$ and $M$ respectively denote the number of $z_c$ and $f_c$ in the $c$-th class.
\end{enumerate}
\end{algorithm}
\end{document}